\documentclass[conference]{ieeeconf}

\usepackage{balance}

\IEEEoverridecommandlockouts
\newcounter{mytempeqncnt}

\newtheorem{definition}{Definition}

\newtheorem{prop}{Proposition}
\newtheorem{thm}{Theorem}

\newtheorem{rmq}{Remark}

\usepackage{amssymb}

\newcommand{\Tordu}[2]{{#1}^{{\theta^{#2}}}}
\newcommand{\Gal}{\operatorname{Gal}}
\newcommand{\rank}{\operatorname{rank}}

\newcommand{\ev}{\operatorname{ev}}

\newcommand{\F}{{\mathbb F}}

\newcommand{\Gab}{\operatorname{Gab}}

\newcommand{\floor}[1]{\lfloor#1\rfloor}

\newcommand{\Q}{\mathbb Q}

\usepackage{amsmath}
\newcommand{\defeq}{\stackrel{\text{def}}{=}}

\newcommand{\MatriceCoins}[4]{
\left( \begin{array}{ccc}
    #1 & \cdots & #2 \\ 
    \vdots & \ddots & \vdots \\ 
    #3 & \cdots & #4 \\ 
  \end{array} \right) 
}

\newcommand{\VecteurLigneCoins}[2]{
\left( #1, \cdots,  #2  \right) 
}

\newcommand{\MatriceGeneraleEch}[3]{
\left( \begin{array}{ccc}
    #1_{1,1} & \cdots & #1_{#3,1} \\ 
    \vdots &  \ddots & \vdots \\ 
    #1_ {1,#2} & \cdots & #1_{#3,#2} \\ 
  \end{array} \right) 
}

\title{\LARGE \bf
Generalization of Gabidulin Codes over Fields of Rational Functions
}

\author{
Daniel Augot
\thanks{$^{1}$ INRIA \& LIX UMR 7161 X-CNRS,
B\^atiment Alan Turing,
Campus de l'\'Ecole polytechnique, 91120 Palaiseau CEDEX, France}
}

\begin{document}
\maketitle

\begin{abstract}
  We transpose the theory of rank metric and Gabidulin codes to the
  case of fields which are not finite fields. The Frobenius
  automorphism is replaced by any element of the Galois group of a
  cyclic algebraic extension of a base field. We use our framework to
  define Gabidulin codes over the field of rational functions using
  algebraic function fields with a cyclic Galois group. This gives a
  linear subspace of matrices whose coefficients are rational
  function, such that the rank of each of this matrix is lower
  bounded, where the rank is comprised in term of linear combination
  with rational functions. We provide two examples based on Kummer and
  Artin-Schreier extensions.The matrices that we obtain may be
  interpreted as generating matrices of convolutional codes.
\end{abstract}

\section{Introduction}
Gabidulin codes are rank-metric codes defined over finite fields using
so-called linearized polynomials~\cite{Gabidulin:1985}. They can be
seen as analogues of Reed-Solomon codes, where polynomials are
replaced by linearized polynomials, and the Hamming distance is
replaced by the rank distance. For Gabidulin codes, the Frobenius
automorphism $x\mapsto x^q$ plays a fundamental
role. In~\cite{ALR:ISIT2013}, the authors generalized this
construction to fields of characteristic zero, where the Frobenius
automorphism does not exists, by considering extensions of number
fields $L/K$, and using a Galois automorphism $\theta$ as the
Frobenius automorphism, and replacing linearized polynomials by
so-called skew polynomials or $\theta$-polynomials. The theory of
Gabidulin transposes nicely, and Maximum Rank Distance  (MRD)
codes can be built, with a decoding algorithm (for the rank distance)
transposed from a simple decoding of Reed-Solomon codes.

In this paper, we use the general abstract framework
of~\cite{ALR:ISIT2013} in the case of the base field being the field
of rational functions (over a finite field). First we briefly recall
facts from~\cite{ALR:ISIT2013}, in the case of a cyclic Galois
extension whose Galois group is generated by $\theta$:
$\theta$-polynomials, rank metric, decoding. We give a construction
with Kummer extensions, a more precise example over the ground field
$\F_8$, and another example with an Artin-Schreier extension over
$\F_5$.

\section{$\theta$-polynomials}
\label{Section:ThetaPoly}
In all the paper, we consider an algebraic field extension $K
\hookrightarrow L$ with finite degree $n$, and an automorphism
$\theta$ in the Galois group $\Gal( K \hookrightarrow L )$, of order
$n=\Gal( K \hookrightarrow L )$.  Given $x \in L$, we use
the notation $x^{\theta^i}$ for $\theta^i(x)$.
%MODIF pour expliquer la notation:
In the finite field case, when $\theta$ is the Frobenius automorphism
$x \mapsto x^q$, $\Tordu{x}{i}=x^{q^i}$, and the similarity is nicely
reflected in this notation.  We define \emph{$\theta$-polynomials},
which are a special case of \emph{skew polynomials}, namely, when
there is no derivation.

\begin{definition} A $\theta$-polynomial is a finite summation of the
  form $\sum_{i\geq0} p_i \Tordu{Z}{i}$, with $p_i \in L$.  The greatest
  integer $i < \infty$ such that $p_i \ne 0$ is called its
  $\theta$-degree, and is denoted by $\deg_{\theta}(P)$.
\end{definition}
We denote the set of $\theta$-polynomials by $L[Z ; \theta]$. This is
a $L$-vector space, which is also a non commutative algebra, using the
skew product:
\[
\left(\sum p_i\Tordu{Z}{i}\right) \cdot \left(\sum
  p_j\Tordu{Z}{j}\right) = \sum_{i,j} p_i\, \Tordu{q_j}{i}
\Tordu{Z}{i+j}.
\]
An evaluation map can also be defined, for $P\in L[Z;\theta]$, and $g\in L$:
\[
\ev(P,g)=P(g)=\sum p_i\Tordu{g}{i}.
\]
The following is well known.
\begin{prop}[\cite{Ore:1932}]
  The set of $\theta$-polynomials $(L[Z ; \theta], + , \cdot)$ is a
  non-commutative integral domain, with unity $\Tordu{Z}{0}=Z$. 
  It is also a left and right Euclidean ring.
\end{prop}
We define the root-space of a $\theta$-polynomial $P(Z)$ to be the set
of $x\in L$ such that $P(x)=0$.  Then we have:
\begin{thm}\label{theo:root}
  The dimension of the root-space of a $\theta$-polynomial is less
  than or equal to its $\theta$-degree.
\end{thm}
\begin{thm} 
\label{Thm:Lagrange}
Let
${V}$ be an $s$-dimensional $K$-subspace of $L$. Then there
exists a unique monic $\theta$-polynomial $P_{{V}}$ with
$\theta$-degree $s$ such that
 \[
  P_{{V}}(x) = 0 \text{ \ for all } x \in {V}.
\]
\end{thm}
See~\cite{ALR:ISIT2013} for proofs of these two Theorems.

\section{Rank Metric and $\theta$-codes}
In this section we recall the definition of the rank weight. All the
proofs are to be found in~\cite{ALR:ISIT2013}.  The codes we are going
to define have codewords $c\in L^n$. We note ${B}=(b_1,\ldots,b_m)$ a
fixed $K$-basis of $L$.  Let $c =\VecteurLigneCoins{c_1}{c_n} \in
L^n$. We define
\[ M_c \defeq \MatriceGeneraleEch{c}{n}{n},
\] where $c_i= \sum_{j=1}^{n} c_{i,j} b_j$. We then define the rank
weight which is related to $K$-linear independence:
\begin{definition}The \emph{rank weight} is defined by 
\[w(c) \defeq \rank_K \left( c_{{B}} \right), \quad \text{for all }c \in L^n.\]
\end{definition}
It is easy to see that the $w$ provides a distance defined by $ d(c_1,c_2)
\defeq w(c_1-c_2)$.  This definition is a generalization of rank metric as
defined in Gabidulin~\cite{Gabidulin:1985}. In~\cite{ALR:ISIT2013}, we
provided four equivalent definitions of the rank metric, a convenient
one being $w(c) \defeq \deg_{\theta}(\min(I_c))$, where $\min(I_c)$ is
the right generator of the ideal
\[ I_c = \left\{ P \in L[Z ; \theta] : P(c_i)=0, \;
i=1,\ldots,n\right\}.
\] 
We also define the generalization of Gabidulin codes.
\begin{definition}
  Let $g=\VecteurLigneCoins{g_1}{g_n} \in L^n$, be $K$-linearly
  independent elements of $L$.  The generalized Gabidulin code, with
  dimension $k$ and length $n$, denoted $\Gab_{\theta,k}(g)$, as a
  $L$-subspace of $L^N$, is $L$-generated by the matrix
\[
G \defeq
\MatriceCoins{\Tordu{g_1}{0}}{\Tordu{g_N}{0}}{\Tordu{g_1}{k-1}}{\Tordu{g_N}{k-1}}.
\]
\end{definition}
Using the evaluation map:
\[
\ev_g:\begin{array}[t]{rcl}
L[Z;\theta]&\rightarrow &L^n\\
P(Z)&\mapsto&\ev_g(P)=\VecteurLigneCoins{P(g_1)}{P(g_n)} 
\end{array}
\]
we may also define the code as  an evaluation code:
\[
\Gab_{\theta,k}(g)=\left\{\ev_g(P):\ P\in L[Z;\theta],\ \deg_\theta P<k\right\}.
\]

For $k \leq n$, the dimension of $\Gab_{\theta,k}(g)$ is indeed $k$.

\begin{prop}[Singleton bound]
  Let ${C}$ be any $[n,k,d]_L$ code for the rank distance. Then $d\le
  n-k+1$.
\end{prop} 
An optimal code satisfying the property that $d=n-k+1$ is called a 
Maximum Rank Distance (MRD) code.
\begin{thm}
The generalized Gabidulin $\Gab_{\theta,k}(g)$ is an MRD code.
\end{thm}
We also briefly recall how to decode these codes. Actually, any
decoding algorithm of Gabidulin codes may be transformed in a decoding
algorithm for our codes, using $\theta$ in place of the Frobenius map:
$x\mapsto x^q$. We present a high level view of the decoding algorithm,
inspired from Gemmel and Sudan's presentation of the algorithm of
Welch-Berlekamp~\cite{Gemmell-Sudan:IPL1992}, but relevant faster
algorithms can be found in~\cite{Loidreau:2005} or more recently
in~\cite{WAS:DCC2013}.  

Consider a vector $y=\VecteurLigneCoins{y_1}{y_n} \in L^n$ such that
there exists $e=\VecteurLigneCoins{e_1}{e_n},\
c=\VecteurLigneCoins{c_1}{c_n}\in L^n$ such that
\begin{IEEEeqnarray*}{rcl}\label{decodingsituation} y = c + e,\\
c \in \Gab_{\theta,k}(g),\\
w(e)\leq \floor{(n-k)/2}.
\end{IEEEeqnarray*}
Write $t=\floor{(n-k)/2}$. We define the following series of problems
related to this situation.
\begin{definition}[Decoding]
Given $y\in L^n$, find, if it exists, a pair $(f,e)$ such that
$y_i=f(g_i)+e_i$, $i=1,\ldots,n$ ; $w(e) \leq t$; $\deg_{\theta}(f) < k$.
%MODIF
%En fait, on ne s'interesse pas a retrouver C, ce qui nous interesse, ce sont uniquement les coefs de f.
%non seulement C ne ne permet pas de retrouver l'information de dÃ©part (on a seulement otÃ© l'erreur et il reste Ã  faire une interpolation "classique"),
%mais en plus on utilise f pour le calculer.
%Given $Y\in L^n$, find, if it exists, a pair $(C,E)$ such that
% $C \in \Gab_{\theta,k}(g)$, $w(E) \leq t$, $Y=C+E$.
\end{definition}
\begin{definition}[Nonlinear reconstruction]
  Given $y\in L^n$, find, if it exists, a pair of $\theta$-polynomials
  $(V,f)$ such that $\deg_{\theta}(V) \leq t$ ; $V \neq 0$ ; 
  $\deg_{\theta}(f) < k$ ;  $ V(y_i) = V(f(g_i))$, $i=1,\ldots,n$.
\end{definition}
Note that this problem gives rise to quadratic equations, considering
as indeterminates the coefficients of the unknowns $(V,f)$ over the
basis ${B}$. We thus consider a linear version of the system.
\begin{definition}[Linearized reconstruction]
  Given $Y\in L^n$, find, if it exists, a pair of $\theta$-polynomials
  $(W,N)$ such that
 $\deg_{\theta}(W) \leq t$ ; $W \neq 0$ ;
 $\deg_{\theta}(N) < k+t$ ;
 $W(y_i) = N(g_i)$, $i=1,\ldots,n$.
\end{definition}
When we have unique decoding, i.e.\ when the weight of the error $e$
is less than or equal to $\floor{(n-k)/2}$, we have the following relations
between the solutions of these problems.
 \begin{prop}
   If $t\leq (n-k)/2$, and if there is a solution to \emph{nonlinear
     reconstruction}, then any solution of \emph{Linear
     reconstruction} gives a solution to \emph{nonlinear
     reconstruction}.
\end{prop}
The solution $f$ can be found by dividing $N$ by $W$.
\begin{rmq}
  The number of arithmetic operations used in this method is easily
  seen to be of $O(n^3)$, using for instance Gaussian elimination for
  solving the linear system. However, since the system is highly
  structured, a better algorithm exists~\cite{Loidreau:2005} whose
  complexity is $O(n^2)$.  This does not reflect the bit-complexity,
  only the arithmetic complexity.
\end{rmq}

\begin{figure*}[!t]
% ensure that we have normalsize text
\normalsize
% Store the current equation number.
\setcounter{mytempeqncnt}{\value{equation}}
% Set the equation number to one less than the one
% desired for the first equation here.
% The value here will have to changed if equations
% are added or removed prior to the place these
% equations are referenced in the main text.
\setcounter{equation}{0}
\begin{equation}
\label{eqn:message}
m^T=\left(\begin{array}{c}
    \frac{\beta^3x + \beta^{10}}{x + \beta^5}y^4 + \frac{\beta^5x + \beta^2}{x + \beta^4}y^3 + \frac{\beta^6x +
        \beta^{13}}{x + \beta^3}y^2 + \frac{\beta^{10}x + \beta^6}{x + \beta^9}y + \frac{\beta x + \beta^{12}}{x + 1}\\
    \frac{\beta^9x + \beta^{14}}{x + \beta^6}y^4 + \frac{\beta^6x + \beta}{x + \beta^4}y^3 + \frac{\beta^{14}x +
        \beta^{13}}{x + \beta^3}y^2 + \frac{\beta^8x + \beta^7}{x + \beta^{12}}y + \frac{\beta^{11}x + \beta^{11}}{x +
        \beta}\\
    \frac{\beta^4x + \beta^{11}}{x + \beta^5}y^4 + \frac{\beta^6x + \beta^{10}}{x + \beta^{11}}y^3 + \frac{\beta^5x +
        \beta^{11}}xy^2 + \frac{\beta^8x + \beta^6}{x + \beta^7}y + \frac{\beta x + \beta^{12}}{x + \beta^6}
\end{array}
\right)
\end{equation}
\begin{equation}
\label{eqn:ec}
c^T=\left(\begin{array}{c}
    \frac{x^2 + \beta^2x + \beta^8}{x^2 + \beta^9x + \beta^{11}}y^4 + \frac{\beta^5x^2 + \beta^7}{x^2 + \beta^{13}x
        + 1}y^3 + \frac{\beta^4x^2 + \beta^7x + \beta^{14}}{x^2 + \beta^3x}y^2 + \frac{\beta^{10}x^3 +
        \beta^4x^2 + \beta^{13}x + \beta^{13}}{x^3 + \beta^{11}x^2 + \beta^{13}x + \beta^{13}}y + \frac{\beta^{11}x^3
        + \beta^{13}x^2 + \beta^3x + \beta^9}{x^3 + \beta^{12}x^2 + \beta^8x + \beta^7}\\
    \frac{\beta^4x + \beta^5}{x + \beta^{11}}y^4 + \frac{\beta^5x^2 + \beta x+ \beta^5}{x^2 + \beta^3x}y^3 +
        \frac{\beta^{11}x^2 + \beta^2x + 1}{x^3 + \beta^{11}x^2 + \beta^{13}x + \beta^{13}}y^2 + \frac{\beta^{10}x +
        \beta^5}{x^3 + \beta^{12}x^2 + \beta^8x + \beta^7}y + \frac{\beta^7x^2 + \beta^6x}{x^2 + \beta^9x +
        \beta^{11}}\\
    \frac{\beta^{11}x^2 + \beta^{13}x + \beta^{11}}{x^2 + \beta^3x}y^4 + \frac{\beta^3x^3 + \beta x^2 + \beta^2}{x^3
        + \beta^{11}x^2 + \beta^{13}x + \beta^{13}}y^3 + \frac{\beta^7x^3 + \beta^7x^2 + 1}{x^3 +
        \beta^{12}x^2 + \beta^8x + \beta^7}y^2 + \frac{\beta^7x^3 + \beta^4x^2 + \beta^6x}{x^2 + \beta^9x +
        \beta^{11}}y + \frac{x^3 + \beta^3x^2 + \beta^7x}{x^2 + \beta^{13}x + 1}\\
    \frac{\beta^{13}x^3 + \beta^6x^2 + \beta^5x + \beta}{x^3 + \beta^{11}x^2 + \beta^{13}x + \beta^{13}}y^4 +
        \frac{\beta^{10}x^3 + \beta^9x^2 + \beta^4x + \beta^{12}}{x^3 + \beta^{12}x^2 + \beta^8x + \beta^7}y^3 +
        \frac{\beta^7x^3 + \beta^{14}x^2 + \beta^{14}x}{x^2 + \beta^9x + \beta^{11}}y^2 + \frac{\beta^{13}x^3 +
        \beta^{12}x^2 + \beta^8x}{x^2 + \beta^{13}x + 1}y + \frac{\beta^6x^2 + x + \beta^2}{x + \beta^3}\\
    \frac{\beta^6x^2 + \beta^6x + 1}{x^3 + \beta^{12}x^2 + \beta^8x + \beta^7}y^4 + \frac{\beta^{14}x^3 +
        \beta^3x^2 + \beta^{11}x}{x^2 + \beta^9x + \beta^{11}}y^3 + \frac{\beta^{12}x^3 + \beta x^2 +
        \beta x}{x^2 + \beta^{13}x + 1}y^2 + \frac{\beta^7x^2 + \beta^3x + \beta^8}{x + \beta^3}y +
        \frac{\beta^8x^4 + \beta^4x^2 + \beta^{13}x}{x^3 + \beta^{11}x^2 + \beta^{13}x + \beta^{13}}
\end{array}
\right)
\end{equation}
\begin{equation}\label{eq:matc}
M_c=\left(\begin{array}{ccccc}
        \frac{\beta^{11}x^3 + \beta^{13}x^2 + \beta^3x + \beta^9}{x^3 + \beta^{12}x^2 + \beta^8x + \beta^7}&
        \frac{\beta^{10}x^3 + \beta^4x^2 + \beta^{13}x + \beta^{13}}{x^3 + \beta^{11}x^2 + \beta^{13}x + \beta^{13}}&
        \frac{\beta^4x^2 + \beta^7x + \beta^{14}}{x^2 + \beta^3x}&
        \frac{\beta^5x^2 + \beta^7}{x^2 + \beta^{13}x + 1}&
        \frac{x^2 + \beta^2x + \beta^8}{x^2 + \beta^9x + \beta^{11}}
\\
        \frac{\beta^7x^2 + \beta^6x}{x^2 + \beta^9x + \beta^{11}}&
        \frac{\beta^{10}x + \beta^5}{x^3 + \beta^{12}x^2 + \beta^8x + \beta^7}&
        \frac{\beta^{11}x^2 + \beta^2x + 1}{x^3 + \beta^{11}x^2 + \beta^{13}x + \beta^{13}}&
        \frac{\beta^5x^2 + \beta x + \beta^5}{x^2 + \beta^3x}&
        \frac{\beta^4x + \beta^5}{x + \beta^{11}}
\\
        \frac{x^3 + \beta^3x^2 + \beta^7x}{x^2 + \beta^{13}x + 1}&
        \frac{\beta^7x^3 + \beta^4x^2 + \beta^6x}{x^2 + \beta^9x + \beta^{11}}&
        \frac{\beta^7x^3 + \beta^7x^2 + 1}{x^3 + \beta^{12}x^2 + \beta^8x + \beta^7}&
        \frac{\beta^3x^3 + \beta x^2 + \beta^2}{x^3 + \beta^{11}x^2 + \beta^{13}x + \beta^{13}}&
        \frac{\beta^{11}x^2 + \beta^{13}x + \beta^{11}}{x^2 + \beta^3x}
\\
        \frac{\beta^6x^2 + x + \beta^2}{x + \beta^3}&
        \frac{\beta^{13}x^3 + \beta^{12}x^2 + \beta^8x}{x^2 + \beta^{13}x + 1}&
        \frac{\beta^7x^3 + \beta^{14}x^2 + \beta^{14}x}{x^2 + \beta^9x + \beta^{11}}&
        \frac{\beta^{10}x^3 + \beta^9x^2 + \beta^4x + \beta^{12}}{x^3 + \beta^{12}x^2 + \beta^8x + \beta^7}&
        \frac{\beta^{13}x^3 + \beta^6x^2 + \beta^5x + \beta}{x^3 + \beta^{11}x^2 + \beta^{13}x + \beta^{13}}
\\
        \frac{\beta^8x^4 + \beta^4x^2 + \beta^{13}x}{x^3 + \beta^{11}x^2 + \beta^{13}x + \beta^{13}}&
        \frac{\beta^7x^2 + \beta^3x + \beta^8}{x + \beta^3}&
        \frac{\beta^{12}x^3 + \beta x^2 + \beta x}{x^2 + \beta^{13}x + 1}&
        \frac{\beta^{14}x^3 + \beta^3x^2 + \beta^{11}x}{x^2 + \beta^9x + \beta^{11}}&
        \frac{\beta^6x^2 + \beta^6x + 1}{x^3 + \beta^{12}x^2 + \beta^8x + \beta^7}
\end{array}
\right)
\end{equation}
% Restore the current equation number.
\setcounter{equation}{\value{mytempeqncnt}}
% IEEE uses as a separator
\hrulefill
% The spacer can be tweaked to stop underfull vboxes.
\vspace*{4pt}
\end{figure*}

\section{Kummer extensions of function fields} 
We now use the previous theory when the field $K$ is a function field
on one variable, the simplest case being $K=k(x)$ the field of
rational functions over a base field $k$. We need to build cyclic
extensions of $k(x)$. 
An standard way of constructing a cyclic extension is to consider a Kummer
extension. The ground field is the field of rational functions, which
then extended by adding a $n$-th root of some element $u\in k(x)$.

We refer the reader to Stichtenoth's book~\cite{Stichtenoth:AFFC1993}
for the theory of algebraic function fields.

For simplicity, we consider the finite field case, when $k=\F_q$, for
some prime power $q$, and $k$ is containing an $n$-root of unity
$\alpha$, for $n$ dividing $q-1$. Note that we can also deal with
fields of characteristic zero like $\Q$, but we may have to extend
them by adjoining $n$-th roots of unity, see~\cite{ALR:ISIT2013}.
Then $K=k(x)$ is the field of rational functions, and for $u\in K$
such $u\neq w^d$, for all $d|n$ and $w\in K$, we can build the field
$L=K[y]$, where $y$ is a root of $Y^n-u=0$. Then $L$ is a cyclic
extension of $K$ of degree $n$, with basis
\[
B=g=\left(1,y,\dots,y^{n-1}\right)
\]
and whose Galois group is generated by $\theta:y\mapsto \alpha y$.  We
can use the previous general framework for building a $\theta$-code
$\Gab_{\theta,k}(g)$ for all $1\leq k\leq n$. A generating matrix is
\[
G \defeq
\MatriceCoins{\Tordu{\left(1\right)}{0}}{\Tordu{\left(y^{n-1}\right)}{0}}{\Tordu{\left(1\right)}{k-1}}{\Tordu{\left(y^{n-1}\right)}{k-1}}.
\]
Then, this matrix defines an MRD code over $L$. Its codewords are of
the form $c=(c_1,\dots,c_n)=(m_1,\dots,m_k)\cdot G$, $c_i,m_i\in L$. Using the basis
$B=\left(1,y,\dots,y^{n-1}\right)$, a codeword can be seen as a matrix
\[
M_c=\MatriceCoins{c_{11}}{c_{1n}}{c_{n1}}{c_{nn}}
\]
where the $c_{ij}$'s are $K=\F_q(x)$. The construction implies that
for any codeword $\rank M_c\geq n-k+1$, where is the rank is
understood in terms of $\F_q(X)$-linear combinations.
\section{A worked out example}
We set $k=\F_{16}=\F_2[\beta]$, with $\beta^4+\beta+1=0$, and we set
$\alpha=\beta^3$, which is a primitive $5$-th root of unity. Then, a
Kummer extension is constructed by adjoining to $K=\F_{16}(x)$, $y$ a
root of $Y^5-x$, which is an irreducible polynomial, to build
$L=K[y]$. The Galois group $\Gal(L\hookrightarrow K)$ has order 5,
with generator $\theta: y\mapsto \alpha y$.
% By extension, $\theta$ is defined as
% \begin{align*}
% \theta(1)&=1\\
% \theta(y)&=y\\
% \theta(y^2)&=\alpha^2y^2\\
% \theta(y^3)&=\alpha^3y^3\\
% \theta(y^4)&=\alpha^4y^4
% \end{align*}
The matrix of the conjugates of the basis is given by:
\[
\left(\begin{array}{ccccccc}
1&y&y^2&y^3&y^4\\
1&\beta^3y&\beta^6y^2&\beta^9y^3&\beta^{12}y^4\\
1&\beta^6y&\beta^{12}y^2&\beta^3y^3&\beta^9y^4\\
1&\beta^9y&\beta^3y^2&\beta^{12}y^3&\beta^6y^4\\
1&\beta^{12}y&\beta^9y^2&\beta^6y^3&\beta^3y^4\\
\end{array}\right)
\]
Picking the first three rows gives a generating matrix for
a 3 dimensional $\theta$-code:
\[
G=
\left(\begin{array}{ccccccc}
1&y&y^2&y^3&y^4\\
1&\beta^3y&\beta^6y^2&\beta^9y^3&\beta^{12}y^4\\
1&\beta^6y&\beta^{12}y^2&\beta^3y^3&\beta^9y^4\\
\end{array}\right)
\]
We give in Eqs.~\ref{eqn:message}, \ref{eqn:ec}, \ref{eq:matc} an
example of a codeword. A message $m\in L^3$ is shown (in transpose
form) in Eq.~\ref{eqn:message}, then $c=m\cdot G\in L^5$ is computed,
as shown in Eq.~\ref{eqn:ec}. We can expand $c$ in the basis
$1,y,\dots,y^4$ to obtain the matrix $M_c$, (Eq.~\ref{eq:matc}).

Note that we obtain matrices with (unbounded) coefficients in
$K=\F_{16}[x]$, where the function field construction with the Kummer
extension may be discarded.

\begin{figure*}[!t]
% ensure that we have normalsize text
\normalsize
% Store the current equation number.
\setcounter{mytempeqncnt}{\value{equation}}
% Set the equation number to one less than the one
% desired for the first equation here.
% The value here will have to changed if equations
% are added or removed prior to the place these
% equations are referenced in the main text.
\setcounter{equation}{3}
\begin{equation}\label{eqn:artiG}
G=\left(
\begin{array}{ccccc}
1 &y &y^2 &y^3 &y^4\\
1& y+1 &y^2 + 2 y + 1 &y^3 + 3 y^2 + 3 y + 1 &y^4 + 4 y^3 + y^2 + 4 y + 1\\
1&y + 2 &y^2 + 4 y + 4 &y^3 + y^2 + 2 y + 3 &y^4 + 3 y^3 + 4 y^2 + 2 y + 1
\end{array}
\right)
\end{equation}
\begin{equation}\label{eqn:artim}
m=\left(
\begin{array}{c}
    \frac{x + 1}{x + 3}y^4 + \frac1xy^3 + (4x + 4)y^2 + \frac{x + 2}xy + \frac{4x + 1}x,\\
    (3x + 2)y^4 + \frac{4x + 3}xy^3 + \frac1{x + 2}y^2 + (2x + 1)y + 1,\\
    \frac2{x + 1}y^4 + \frac{4x + 4}{x + 2}y^3 + 4y^2 + y + \frac3{x + 1}
\end{array}
\right)
\end{equation}
\begin{equation}\label{eqn:artic}
c^T=\left(
\begin{array}{c}
    \frac{3x^3 + x + 3}{x^2 + 4x + 3}y^4 + \frac{3x^2 + x + 3}{x^2 + 2x}y^3 +
        \frac{4x^2 + x + 2}{x + 2}y^2 + \frac{2x^2 + 3x + 2}xy + \frac{4x + 1}{x^2 +
        x}\\
    \frac{3x^4 + 4x^3 + x + 3}{x^3 + 3x^2 + 2x}y^4 + \frac{4x^3 + 3x^2 + x +
        1}{x^2 + 2x}y^3 + \frac{2x^3 + 4}{x^2 + 2x}y^2 + \frac{x^2 + 2x + 3}{x^2
        + 3x}y + \frac{3x^4 + 2x^2 + 3x + 1}{x^2 + 4x + 3}\\
    \frac{2x^3 + x^2 + 3x + 1}{x^2 + x}y^4 + \frac{2x^2 + 3x + 4}{x + 2}y^3 +
        \frac{2x^4 + 2x^2 + 4x + 1}{x^3 + x}y^2 + \frac{3x^6 + 4x^5 + 3x^4 + 2x^3
        + x^2 + x + 4}{x^4 + x^3 + x^2 + x}y + \frac{x^4 + 2x^2 + 3x + 4}{x^2 +
        3x + 2}\\
    \frac{4x^2 + x + 1}{x^2 + x}y^4 + \frac{4x^4 + x^3 + 2x^2 + 4}{x^3 + x}y^3 +
        \frac{3x^6 + x^5 + 2x^2 + 4x + 4}{x^4 + x^3 + x^2 + x}y^2 + \frac{4x^4 +
        3x^3 + 3x^2 + x + 4}{x^2 + x}y + \frac{3x^3 + x^2 + x + 3}{x + 2}\\
    \frac{4x^5 + x^4 + 2x^3 + 2x + 3}{x^4 + x^3 + x^2 + x}y^4 + \frac{3x^6 + x^4 +
        3x^3 + 2x^2 + x + 2}{x^4 + x^3 + x^2 + x}y^3 + \frac{2x^5 + 2x^4 +
        2x^3 + x^2 + 2x + 4}{x^3 + 3x^2 + 2x}y^2 + \frac{2x^3 + x^2 + 2x}{x
        + 1}y + \frac{4x^4 + 3x + 3}{x^2 + 3x + 2}
\end{array}
\right)
\end{equation}
\begin{equation}\label{eqn:artimat}
M_c=\left(
\begin{array}{ccccc}
        \frac{4x + 1}{x^2 + x}&
        \frac{2x^2 + 3x + 2}x&
        \frac{4x^2 + x + 2}{x + 2}&
        \frac{3x^2 + x + 3}{x^2 + 2x}&
        \frac{3x^3 + x + 3}{x^2 + 4x + 3}
    \\
        \frac{3x^4 + 2x^2 + 3x + 1}{x^2 + 4x + 3}&
        \frac{x^2 + 2x + 3}{x^2 + 3x}&
        \frac{2x^3 + 4}{x^2 + 2x}&
        \frac{4x^3 + 3x^2 + x + 1}{x^2 + 2x}&
        \frac{3x^4 + 4x^3 + x + 3}{x^3 + 3x^2 + 2x}
        \\
        \frac{x^4 + 2x^2 + 3x + 4}{x^2 + 3x + 2}&
        \frac{3x^6 + 4x^5 + 3x^4 + 2x^3 + x^2 + x + 4}{x^4 + x^3 + x^2 + x}&
        \frac{2x^4 + 2x^2 + 4x + 1}{x^3 + x}&
        \frac{2x^2 + 3x + 4}{x + 2}&
        \frac{2x^3 + x^2 + 3x + 1}{x^2 + x}
        \\
        \frac{3x^3 + x^2 + x + 3}{x + 2}&
        \frac{4x^4 + 3x^3 + 3x^2 + x + 4}{x^2 + x}&
        \frac{3x^6 + x^5 + 2x^2 + 4x + 4}{x^4 + x^3 + x^2 + x}&
        \frac{4x^4 + x^3 + 2x^2 + 4}{x^3 + x}&
        \frac{4x^2 + x + 1}{x^2 + x}
        \\
        \frac{4x^4 + 3x + 3}{x^2 + 3x + 2}&
        \frac{2x^3 + x^2 + 2x}{x + 1}&
        \frac{2x^5 + 2x^4 + 2x^3 + x^2 + 2x + 4}{x^3 + 3x^2 + 2x}&
        \frac{3x^6 + x^4 + 3x^3 + 2x^2 + x + 2}{x^4 + x^3 + x^2 + x}&
        \frac{4x^5 + x^4 + 2x^3 + 2x + 3}{x^4 + x^3 + x^2 + x}
\end{array}
\right)
\end{equation}
\setcounter{equation}{\value{mytempeqncnt}}
% IEEE uses as a separator
\hrulefill
% The spacer can be tweaked to stop underfull vboxes.
\vspace*{4pt}
\end{figure*}

\section{Artin-Schreier case}
For completeness, we describe the Artin-Schreier situation, which is
particular to the positive characteristic case. The theory of such
extensions is also described in~\cite{Stichtenoth:AFFC1993}. Assume
that $k$ has characteristic $p$, and consider $K=k(x)$, with an
element $u\in K$ such that
\[
u\neq w^p-w\text{  for all } w \in K.
\]
Then the extension $L=K[y]$, where $y$ is a root of $Y^p-Y=w$ is an
Artin-Schreier extension. Its Galois group is cyclic of order $p$,
whose generator $\theta$ is defined by $\theta(y)=y+1$. Consider as an
example $k=\F_5$, $K=k(x)$, and $L=K[y]$, with $y^5-y=x$. Then
$\theta(y)=y+1$, and we can build a $[5,3,3]_L$ code with generating
matrix $G$ given in Eq.~\ref{eqn:artiG}. We
give in Eqs.~\ref{eqn:artim}, \ref{eqn:artic}, \ref{eqn:artimat} an
example of a codeword. 

% A message $m\in L^3$ is shown (in transpose
% form) in Eq.~\ref{eqn:message}, then $c=m\cdot G\in L^5$ is computed,
% as shown in Eq.~\ref{eqn:ec}. We can expand $c$ in the basis
% $1,y,\dots,y^4$ to obtain the matrix $M_c$, (Eq.~\ref{eq:matc}).

\section{Polynomial matrices}
We briefly mention that in both constructions the basis are integral
bases of $L/K=L/k(x)$. The generating matrices $G$ consist of integral
elements. In that case, we can choose our messages $m\in k[x,y]$
instead of $k(x)[y]$, and the corresponding codewords will also belong
to $k[x,y]$. When  the codewords are expanded as matrices, we
find $n\times n$ matrices with polynomial coefficients.

\section{Conclusion}
We have generalized Gabidulin codes to the field of rational
functional, using cyclic extensions $L$ of $k(x)$, for instance Kummer
extensions, or Artin-Schreier extensions. We can easily find
generating matrices for codes with symbols in $L$. These codewords,
when expanded over $k(x)$ give naturally matrices which have high
rank, where the rank has to understood by considering $k(x)$ linear
combinations of the rows of the matrix. When the Gabidulin code has
dimension $k$, each of these matrices has $k(x)$-rank at least
$n-k+1$, since the codes are Maximum Rank Distance. Given such a
matrix, when its weight, i.e.\ its rank, is $w$, its gives rise to a
rate $w/n$ convolutional codes, using the language of rational
fractions as in~\cite{HuffPless:2003}, replacing $x$ with $D$, the
delay operator. We did not consider the framework of Laurent series as
in~\cite{Roth:ITCT2006}, but we think we can adapt the general theory
to this field.

\balance
\section{Acknowledgments}
We are thankful to V.~Sidorenko for suggesting us to expand the
framework of~\cite{ALR:ISIT2013} to the context of rational function
fields. We also thank Hans-Andrea Loeliger, Emina Soljanin, and Judy
L.~Walker, the organizers of the Dagstuhl ``Coding Theory'' Seminar,
25--30 August 2013, for providing a nice atmosphere for discussing
these topics.

 \bibliographystyle{IEEEtran} \bibliography{mtns}
\end{document}